\documentclass[12pt, a4paper]{scrartcl}

\usepackage[utf8]{inputenc}
\usepackage[T1]{fontenc}
\usepackage[USenglish]{babel}

\usepackage[tracking=true, protrusion=true, expansion=true,final, babel]{microtype}
\usepackage{lmodern}

\usepackage{setspace}

\usepackage[pdfborderstyle={/S/U/W 1}]{hyperref}

\usepackage[version=4]{mhchem} 

\usepackage{amssymb}

\usepackage{graphicx}
\usepackage[labelfont=bf]{caption}
\usepackage{placeins}
\usepackage{booktabs}

\usepackage{lineno}

\usepackage[separate-uncertainty]{siunitx}
\DeclareSIUnit\at{at.\%}
\DeclareSIUnit\unitgravity{\times \, g}
\DeclareSIUnit\ppm{ppm}
\DeclareSIUnit\Molar{\textsc{m}}
\DeclareSIUnit\rpm{rpm}

\usepackage[style=numeric, maxbibnames=99, sorting=none, citestyle=numeric-comp, giveninits=true]{biblatex}
\usepackage{csquotes}

\usepackage{soul} 

\usepackage{authblk}
\usepackage{orcidlink}

\title{Near-Atomic-Scale Compositional Complexity in a 2D Transition Metal Oxide}

\date{}

\author[1,*]{Mathias Krämer \orcidlink{0000-0002-1352-9064}}

\author[2]{Bar Favelukis \orcidlink{0009-0006-6446-0817}}

\author[1]{J. Manoj Prabhakar \orcidlink{0000-0002-1842-0594}}

\author[3]{Aleksander Albrecht \orcidlink{0000-0002-1338-5136}}

\author[2]{Brian A. Rosen \orcidlink{0000-0002-2265-8802}}

\author[2]{Noam Eliaz \orcidlink{0000-0002-1184-4706}}

\author[2]{Maxim Sokol \orcidlink{0000-0001-8588-0585}}

\author[1,4,*]{Baptiste Gault \orcidlink{0000-0002-4934-0458}}

\affil[1]{Max Planck Institute for Sustainable Materials, Max-Planck-Straße 1, 40237 Düsseldorf, Germany}

\affil[2]{Department of Materials Science and Engineering, Tel Aviv University, P.O.B 39040, Ramat Aviv 6997801, Israel}

\affil[3]{Department of Inorganic Chemical Technology and Environment Engineering, Faculty of Chemical Technology and Engineering, West Pomeranian University of Technology in Szczecin, Piastów Ave. 42, Szczecin, 71-065 Poland}

\affil[4]{Univ Rouen Normandie, CNRS, INSA Rouen Normandie, Groupe de Physique des Matériaux, UMR 6634, F-76000 Rouen, France}

\affil[*]{Corresponding authors: m.kraemer@mpi-susmat.de, baptiste.gault1@univ-rouen.fr}

\begin{document}

\maketitle
\onehalfspacing

\clearpage

\section*{Abstract}

2D materials hold transformative promise for next-generation nanoelectronics. However, successfully integrating these materials from laboratory-scale discoveries into real-world devices depends on precisely controlling their properties, which are fundamentally determined by their composition. Detailed characterisation using atom probe tomography of 2D \ce{Ti_{0.87}O2}, a candidate high-$\kappa$ dielectric, reveals deviations from its commonly assumed stoichiometry. Compositional analysis and comparison with the bulk \ce{K_{0.8}[Ti_{1.73}Li_{0.27}]O4} precursor evidences an oxygen deficit indicative of oxygen vacancy formation in the 2D material, as well as the retention of low concentrations of alkali metals that were presumed to be removed during synthesis. Such deviations from stoichiometry indicate a reconstruction mechanism that mitigates the effect of the characteristic, negatively charged vacancies on the titanium sublattice, thereby influencing the local electronic structure and, consequently, functional properties. These findings underscore the importance of a detailed compositional analysis in both understanding and optimizing the extraordinary functional properties of 2D materials, opening pathways to tailored functionalities in next-generation nanoelectronics.

\begin{figure}[!htb]
   \centering
    \includegraphics[width=0.75\textwidth]{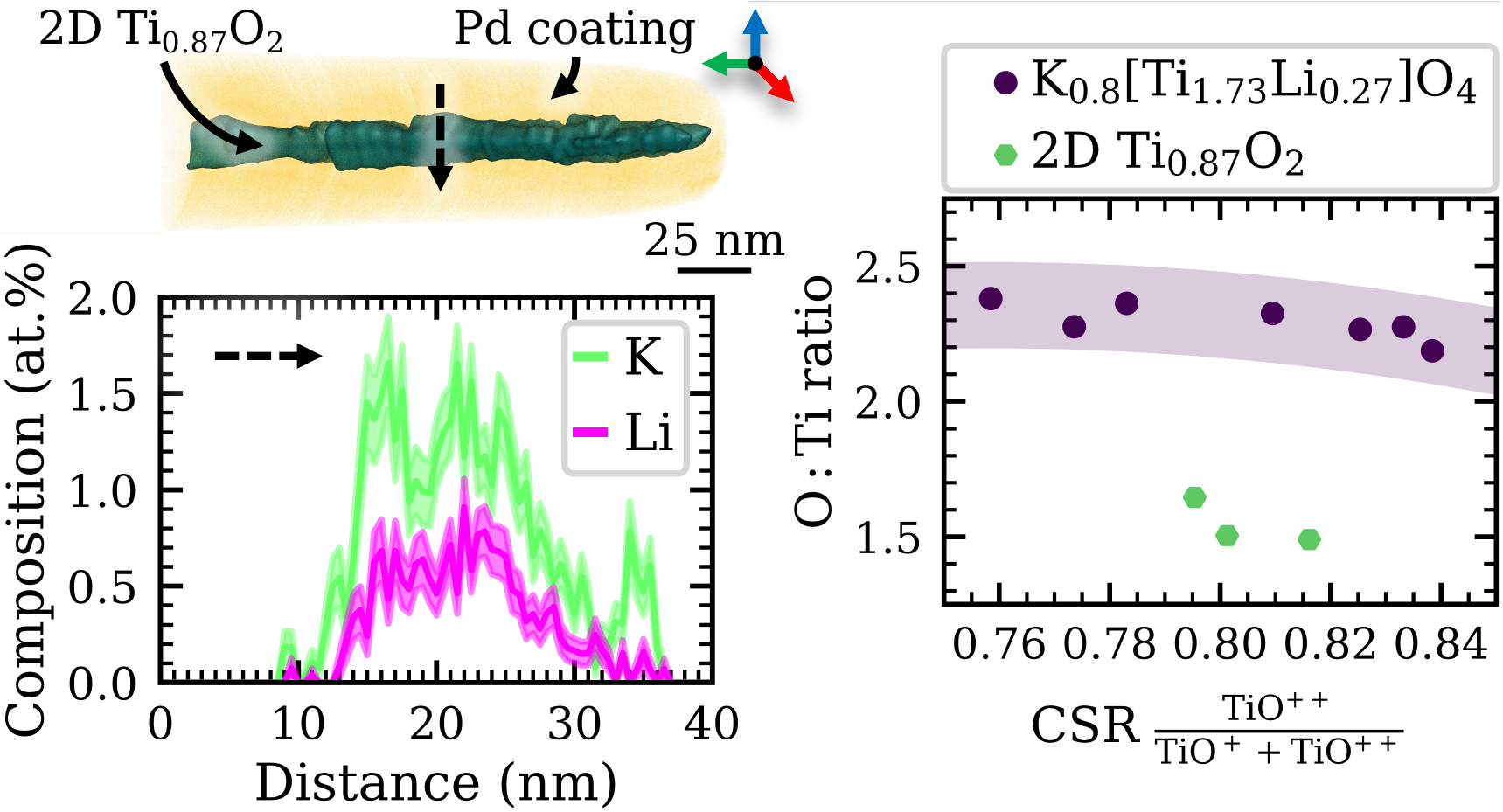}
\end{figure}

\section*{Keywords}

2D Transition Metal Oxides, High-$\kappa$ Dielectric, Nanoelectronics, Atom Probe Tomography, Oxygen Vacancies, Dopants

\clearpage

The continued miniaturization of integrated circuits presents significant technological challenges, particularly concerning device architecture, thereby prompting materials innovation~\cite{2023Cao}. In this context, 2D materials are promising candidates for future micro- and nanoelectronics due to their atomic-scale thickness and wide-ranging functional properties~\cite{2021Das, 2024Zen}. These materials can be readily processed into inks, enabling seamless integration with existing coating and printing technologies, even in the form of complex heterostructures~\cite{2022Pin}. Such heterostructures are therefore also being investigated to build thin-film electronics in flexible or wearable devices~\cite{2014Aki, 2019Lin, 2024Kat}.

Materials under investigation include 2D transition metal oxides (TMOs), which have emerged as possible alternatives to replace dielectrics used to date~\cite{2012Osa, 2019Osa}. Within this group, 2D titanium-based TMOs are particularly interesting due to their remarkable chemical versatility, enabling the synthesis of titanium-deficient non-stoichiometric oxides with tunable properties~\cite{2014Wan}. Specifically, \ce{Ti_{0.87}O2} exhibits a high dielectric constant and a low leakage current density, even at thicknesses below \SI{10}{\nano\metre}, which makes it suitable for use as a high-$\kappa$ dielectric~\cite{2006Osa, 2008Osa, 2009Aka}. Its functionality extends beyond that of a dielectric, with demonstrated applications as an anode material for sodium ion batteries~\cite{2018Xio}, and as a functional interface in alkali metal batteries~\cite{2021Xio, 2023Zha}.

Despite extensive research into their synthesis, the detailed compositions of \ce{Ti_{0.87}O2} and related 2D TMOs have not yet been sufficiently investigated. One of the manufacturing challenges in successfully integrating these materials into next-generation nanoelectronics is achieving precise control over composition and dopants, such as defects and adsorbates~\cite{2022Lem}. Starting with a layered alkali metal TMO crystal, 2D TMOs are synthesised in a two-step cation exchange process, i.e. replacing alkali metal ions with protons~\cite{1982Iza, 1998Sasa}, followed by exfoliation induced by the intercalation of bulky organic ions~\cite{1998Sas}. Questions remain as to whether all alkali metal ions are completely removed during the protonation step, or if they become stabilised at defects, thereby influencing the electronic properties.

High-resolution scanning transmission electron microscopy is a powerful technique for structural and compositional analysis of 2D materials~\cite{2021Odo}, yet identifying light trace elements in the material is very difficult~\cite{2015Sen, 2021Lin}. The characterisation of 2D TMOs is further complicated by beam-induced oxygen vacancy formation, which can cause significant structural and compositional changes, including atomic reconstructions~\cite{2010Wan, 2011Ohw, 2014Ohw, 2022Cho}. Due to the increased interest and demand for the detailed 3D compositional characterisation of nanomaterials, atom probe tomography (APT) has become a well-established complementary technique~\cite{2021Gau}. Particularly in the field of energy materials, APT has contributed to a reconsideration of how elements incorporated during synthesis influence functional properties~\cite{2025Gau}. Even with controlled synthesis conditions, recent studies have revealed the presence of impurities in 2D materials such as 2D \ce{MoS2}~\cite{2020Kim} and \ce{Ti3C2T_{x}} MXenes~\cite{2024Kra_AM}, caused by the unavoidable contamination even with high-purity chemicals used for synthesis. Secondary-ion mass spectrometry has likewise shown that the composition of 2D materials can be more complex than often assumed~\cite{2022Mic}.

Here, we report on a detailed characterisation of the composition of 2D \ce{Ti_{0.87}O2} using APT. By optimizing a previous APT specimen preparation workflow for 2D materials~\cite{2024Kra_MAM}, the in situ sputtering of a chemically inert palladium coating onto the specimen enables reliable quantification of oxygen in the 2D TMO. A comparative analysis with the layered \ce{K_{0.8}[Ti_{1.73}Li_{0.27}]O4} precursor revealed a significant loss of oxygen during synthesis, pointing toward the formation of oxygen vacancies. In addition to the altered oxygen stoichiometry, alkali metals originating from the precursor were also detected in the 2D material. Both the observed oxygen deficiency and the incorporation of alkali metals ions could serve to compensate for the negatively charged titanium vacancies inherent in the non-stoichiometric oxide, thereby influencing defect charge balance and local bonding environments. These findings provide new insights into the more complex composition and defect chemistry of \ce{Ti_{0.87}O2}, and emphasise the need to reassess the composition of 2D TMOs critically.

\section*{Results and Discussion}

\subsection*{Synthesis and Structural Characterisation}

\begin{figure}[!htb]
   \centering
    \includegraphics[width=\textwidth]{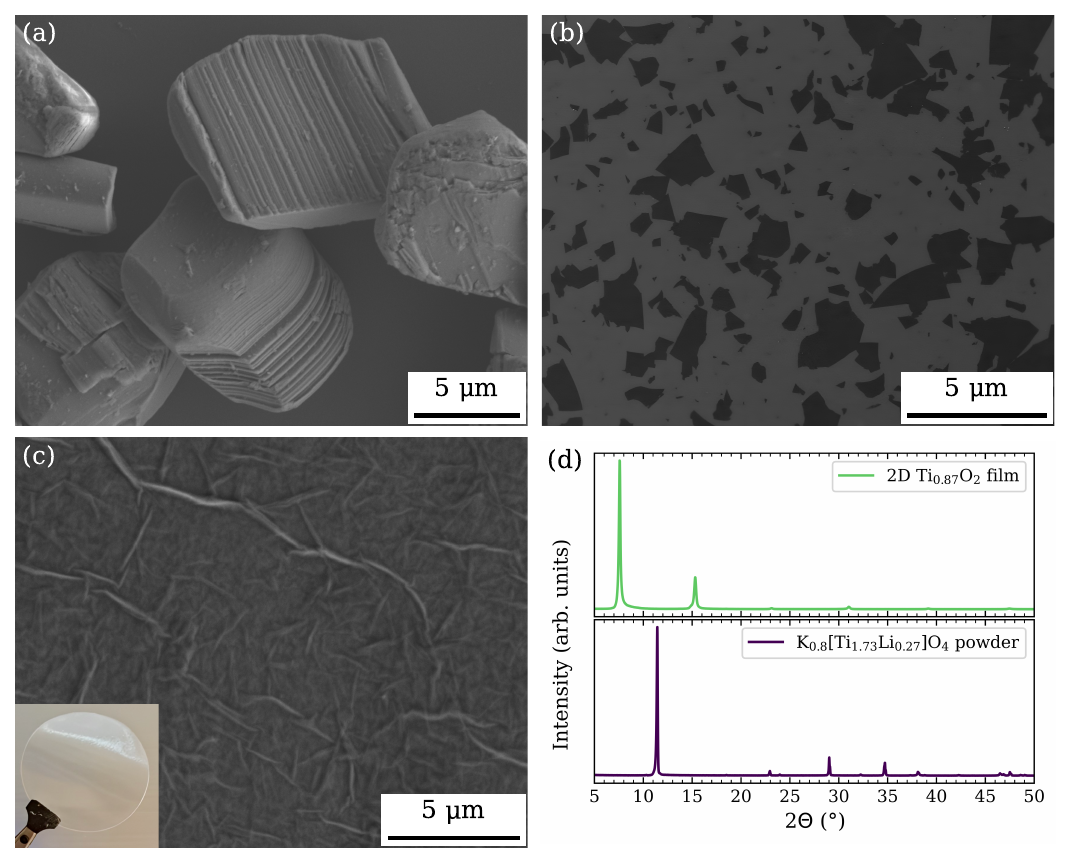}
   \caption{Synthesis of the potassium lithium titanate precursor and derived 2D transition metal oxide. (a) Synthesised \ce{K_{0.8}[Ti_{1.73}Li_{0.27}]O4} powder. (b) Exfoliated 2D \ce{Ti_{0.87}O2} nanosheets. (c) Top-view electron micrograph of the 2D material film, with inset optical photograph highlighting its optical transparency. All scanning electron microscopy images were acquired using the secondary electron imaging mode. (d) X-ray diffraction data for the lepidocrocite-type \ce{K_{0.8}[Ti_{1.73}Li_{0.27}]O4} powder and the 2D \ce{Ti_{0.87}O2} film.}
    \label{fig:Synthesis}
\end{figure}

The potassium lithium titanate precursor was synthesised and exfoliated into 2D \ce{Ti_{0.87}O2} nanosheets via cation exchange, as confirmed by scanning electron microscopy (SEM) and X-ray diffraction (XRD). The characteristic layered structure of the \ce{K_{0.8}[Ti_{1.73}Li_{0.27}]O4} precursor powder is evident in the SEM image shown in \autoref{fig:Synthesis}~(a). Following exfoliation, SEM imaging of a diluted, spin-coated suspension (\autoref{fig:Synthesis}~(b)) reveals thin, sheet-like morphologies with lateral dimensions characteristic of monolayer nanosheets. For further characterisation, the suspension containing the nanosheets was vacuum-filtered into a macroscopic, free-standing film. A top-view SEM image and an optical photograph are provided in \autoref{fig:Synthesis}~(c). XRD analysis confirms the layered lepidocrocite-type titanate structure of the precursor powder, and reveals sharp (020) and weaker (040) reflections in the 2D TMO film, indicating a reassembly of the nanosheets into a well-ordered stack. Both diffraction patterns are presented in \autoref{fig:Synthesis}~(d).

\subsection*{Revisiting Cation Oxidation State in 2D \ce{Ti_{0.87}O2}}

In discussions regarding the dielectric performance of 2D \ce{Ti_{0.87}O2}, the absence of oxygen vacancies is considered as a key contributing factor~\cite{2012Osa, 2019Osa}. Oxygen vacancies can significantly influence the electronic properties of TMOs by introducing donor-like defect states, generally described as n-type doping~\cite{2012Gre}. These point defects act as electron traps, giving rise to leakage currents that compromise the suitability of these materials as dielectrics by degrading their insulating behaviour~\cite{2020Gun}.

Recent approaches, such as re-oxidation and p-type doping, have been employed to reduce or prevent the formation of oxygen vacancies, thereby improving dielectric integrity~\cite{2000Tem, 2005Lee, 2009Pul}. In the case of \ce{Ti_{0.87}O2}, it has been proposed that the deliberate removal of lithium from the \ce{K_{0.8}[Ti_{1.73}Li_{0.27}]O4} precursor during synthesis generates negatively charged vacancies within the titanium sublattice. These titanium vacancies inhibit the formation of oxygen vacancies~\cite{2011Wan}, thereby maintaining dielectric functionality.

While X-ray photoelectron spectroscopy (XPS) enables detection of oxidation state variations that may imply oxygen vacancy formation, data interpretation remains particularly complex for TMO nanomaterials~\cite{2024Lu, 2024Wan}. In \ce{Ti_{0.87}O2}, this issue is further complicated by a limited understanding of how existing titanium vacancies influence the local electronic structure, and consequently the spectroscopic data. This leaves a critical gap in the interpretation of experimental XPS data for 2D \ce{Ti_{0.87}O2}. For instance, despite the significant alterations to the local structural and coordination environments induced by the cation vacancies, XPS spectra of 2D \ce{Ti_{0.87}O2}~\cite{2008Osa, 2021Xio, 2025Ryu, 2023Zha} remain nearly indistinguishable from those of stoichiometric \ce{TiO2}~\cite{1996Die, 2010Bie}.

\begin{figure}[!htb]
   \centering
    \includegraphics[width=\textwidth]{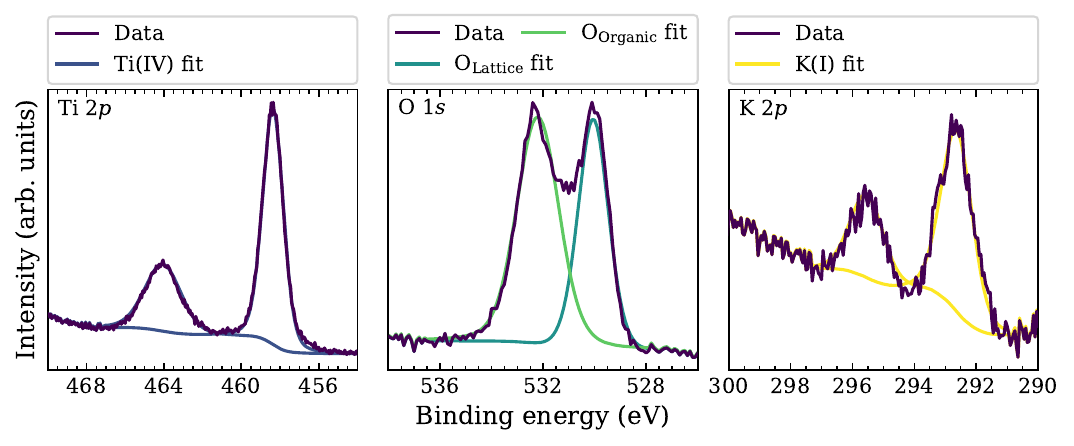}
   \caption{High-resolution X-ray photoelectron \ce{Ti} $2p$, \ce{O} $1s$, and \ce{K} $2p$ core level spectra for the 2D \ce{Ti_{0.87}O2} film.}
    \label{fig:XPS}
\end{figure}

The XPS data of the \ce{Ti} $2p$ core level spectrum in \autoref{fig:XPS} is consistent with previous studies, exhibiting features associated solely to the \ce{Ti(IV)} oxidation state. However, density functional theory calculations indicate that the charge distribution even in \ce{TiO2} may be more complex than commonly assumed, with titanium sites potentially capable of undergoing further oxidation~\cite{2017Koc}. Such a scenario would be expected to produce a shoulder at higher binding energies in the \ce{Ti} $2p$ core level spectrum, and a shoulder at lower binding energies in the \ce{O} $1s$ core level spectrum, due to the cation deficiency and hence different oxidation of the surrounding elements. Yet no such feature is observed here. Interestingly, potassium was detected by XPS, corroborating earlier findings by Zhang et al.~\cite{2025Zha}. However, due to the limited spatial resolution of the technique, its precise distribution within the material remains unresolved.

Being a non-stoichiometric oxide, 2D \ce{Ti_{0.87}O2} contains a relatively high concentration of \ce{Ti} vacancies. Contrast variations in high-resolution transmission electron microscopy combined with image simulations, indicate that titanium vacancies are typically associated with neighbouring oxygen vacancies~\cite{2013Ohw}. These oxygen vacancies may be caused by electron-beam-induced reduction during data acquisition~\cite{2011Ohw}, yet this does not preclude the possibility of such vacancies being present in the material prior to electron imaging. The formation of titanium vacancies during synthesis may trigger a reconstruction mechanism involving the formation of oxygen vacancies, thereby maintaining charge neutrality and yielding a bonding environment closely resembling that of stoichiometric \ce{TiO2}. Such a reconstruction mechanism following synthesis would lead to compositional deviations in the oxygen to titanium ratio compared to the precursor material, which were systematically investigated using APT.

\subsection*{Compositional Deviations in 2D \ce{Ti_{0.87}O2} from the \ce{K_{0.8}[Ti_{1.73}Li_{0.27}]O4} Precursor}

APT was employed to characterise the composition of the 2D material \ce{Ti_{0.87}O2} following synthesis through 3D compositional mapping at the near-atomic-scale. Recent advances in specimen preparation have overcome the earlier limitations of APT to bulk materials, and now enable its more routine application to nanomaterials~\cite{2018Kim, 2025Nal} and, in particular, 2D materials~\cite{2024Kra_MAM}. For these materials, a revised workflow based on the common focused ion beam specimen preparation technique enables the analysis of 2D material films, such as shown in \autoref{fig:Synthesis} (c). An important step in the revised workflow is a metallic coating deposited in situ, which helps to mechanically stabilize the fragile APT specimens of the 2D material film. Here, chemically inert palladium is introduced to minimize oxygen incorporation during the coating procedure, which can result from surface passivation, to ensure precise quantification of the oxygen content in the TMO (see Supporting Information for more details).

While specimen preparation for nanomaterials presents its own set of challenges, APT analysis of oxides is complicated by compositional biases arising from the complex field evaporation behaviour of these materials. Studies on various oxides have demonstrated that evaporating these materials under high electrostatic field conditions is essential to minimize compositional biases~\cite{2013Dev, 2013Kir, 2014Man, 2015San, 2023Hun, 2024Kim}. The bias in measured oxygen content is attributed to the formation of neutral \ce{O2} forming from the dissociation of oxygen-bearing molecular ions. Although the neutral \ce{O2} molecule carries momentum, it is not accelerated sufficiently to be detected by the particle-detector~\cite{2016Gau}. However, it is sufficiently accelerated to strongly limit ionization near the specimen apex, particularly under low electrostatic field conditions~\cite{2017Zan}.

\begin{figure}[!htb]
    \centering
    \includegraphics[width=0.5\textwidth]{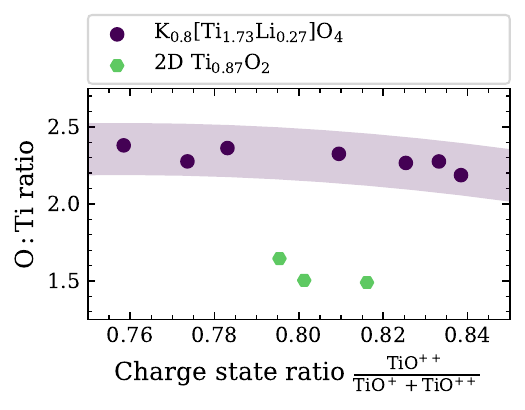}
    
    \caption{Influence of the electrostatic field conditions on the measured oxygen to titanium ratio in bulk \ce{K_{0.8}[Ti_{1.73}Li_{0.27}]O4} and 2D \ce{Ti_{0.87}O2}, as inferred from the \ce{TiO++} to \ce{TiO+} charge state ratio. The fitted line for the bulk \ce{K_{0.8}[Ti_{1.73}Li_{0.27}]O4} data points is intended solely as a guide to illustrate the calibration curve.}
    \label{fig:CSR}
\end{figure}

To determine the influence of the electrostatic field on the composition, the bulk powder with a known stoichiometry was analysed under increasing electrostatic field conditions, i.e. by varying the laser pulse energy while maintaining a constant detection rate. The electrostatic field strength at the specimen apex can be estimated using the charge state ratio (CSR) of a selected atomic or ionic species~\cite{1982Kin, 2024Kim}. By plotting the measured composition against the CSR, as in \autoref{fig:CSR}, a calibration curve is constructed that correlates composition with electrostatic field strength, which is then used to estimate the composition of the 2D TMO. Here, the ion species \ce{TiO+} and \ce{TiO++} were used for the CSR, using only the mass peaks containing the isotopes \ce{^{46}Ti} and \ce{^{47}Ti}, as \ce{^{48}TiO++} overlaps with \ce{O2+} in the mass spectrum at \SI{32}{\dalton}.

The calibration curve in \autoref{fig:CSR} follows the trend reported by Verberne et al. for rutile \ce{TiO2}~\cite{2019Ver}, as a decrease in the \ce{TiO++} to \ce{TiO+} CSR is likewise observed for \ce{K_{0.8}[Ti_{1.73}Li_{0.27}]O4} at lower laser pulse energies. While this behaviour contradicts post-ionisation theory~\cite{1982Kin}, it has also been observed in the \ce{Fe++}/\ce{Fe+} CSR in magnetite (\ce{Fe3O4}), indicating that increasing laser energies drive direct evaporation into higher charge states~\cite{2014Sch}.

A comparison of the measured composition of the 2D material with the calibration curve in \autoref{fig:CSR} of the precursor material suggests a significant compositional change of the 2D TMO following synthesis, rather than a compositional bias. Considering the \ce{TiO2}-like oxidation states of O and Ti identified by XPS, the observed oxygen deficiency relative to the precursor material suggests the formation of oxygen vacancies in the 2D TMO following synthesis and processing. As previously discussed, this points to a reconstruction mechanism that leads to a thermodynamically more stable bonding environment, closely resembling that of stoichiometric \ce{TiO2}. 

Such oxygen loss is reminiscent of processes observed in battery materials, where irreversible oxygen loss is one of the critical degradation mechanism in layered TMO cathode materials for lithium-ion batteries~\cite{2022Zha}, and recently also observed in an inorganic solid electrolyte during cell operation~\cite{2025Cha}. For TMO battery materials, severe delithiation destabilizes the layered structure, triggering a phase transition to a rock-salt structure driven by oxygen release and cation migration~\cite{2014Jun, 2014Lin}. During the synthesis of 2D \ce{Ti_{0.87}O2}, alkali metal ions are replaced by protons through a diffusion-controlled cation exchange reaction, which fundamentally differs from an electrochemical delithiation. However, certain oxygen sites in disordered materials can possess negative oxygen vacancy formation energies, facilitating spontaneous oxygen release even for moderate delithiation~\cite{2025Cse}. Oxygen site stability appears to be closely linked to the local cation coordination environment~\cite{2025Jeo}. In \ce{K_{0.8}[Ti_{1.73}Li_{0.27}]O4}, such thermodynamically unstable oxygen sites may arise from cation disorder caused by lithium occupying titanium sublattice sites, thereby facilitating the formation of oxygen vacancies during the cation exchange reaction. 

\subsection*{Stabilisation of Alkali Metals in 2D \ce{Ti_{0.87}O2}}

\begin{figure}[!htb]
    \centering
    \includegraphics[width=\textwidth]{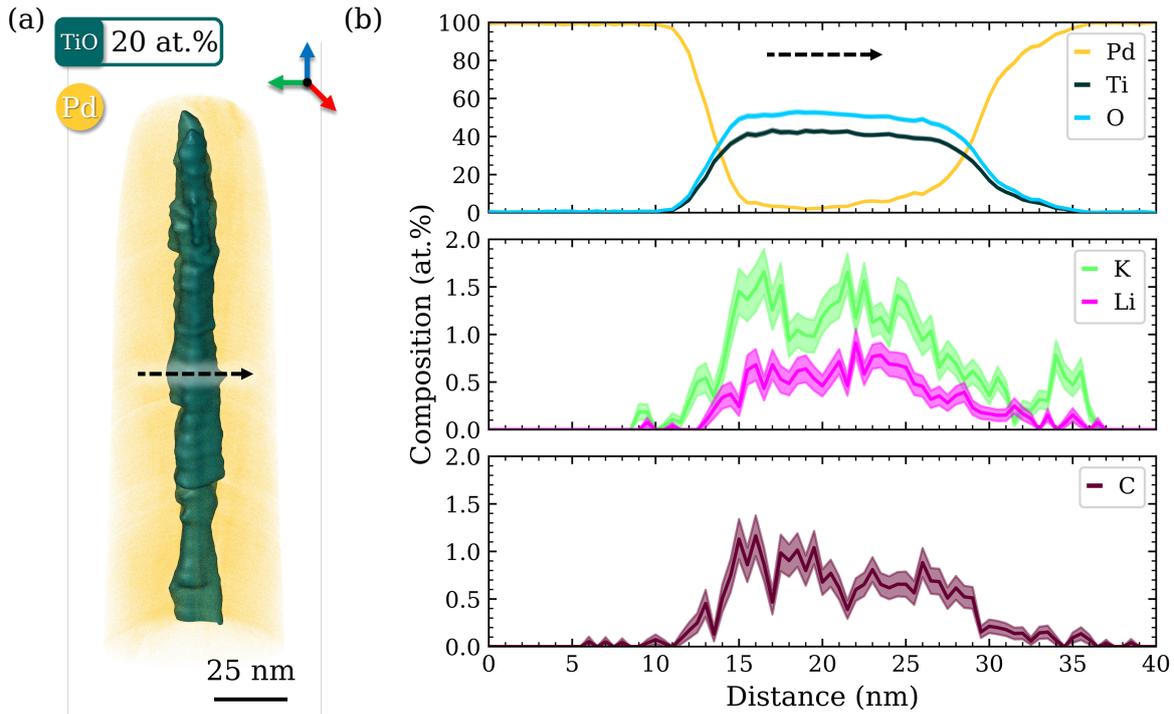} 
    \caption{APT analysis of the 2D \ce{Ti_{0.87}O2} material coated in situ with palladium. (a) Reconstructed 3D atom map. (b) 1D compositional profile ($\varnothing$\SI{10}{\nm}~x~\SI{40}{\nm}) across the region of interest as indicated in (a). Errors are estimated according to counting statistics.}
    \label{fig:Alkali}
\end{figure}

The reconstructed 3D atom map in \autoref{fig:Alkali}~(a) shows the TMO nanosheet stack, highlighted by the dark cyan iso-composition surface, uniformly coated with palladium. A 1D compositional profile in \autoref{fig:Alkali}~(b) through the material of interest reveals the presence of alkali metals originating from the bulk \ce{K_{0.8}[Ti_{1.73}Li_{0.27}]O4} precursor. Lithium and potassium remain in the 2D material at concentrations of \SI[separate-uncertainty = true]{0.39(1)}{\at} and \SI{1.16(2)}{\at}, respectively. As previously noted, potassium was also detected by XPS. Carbon was quantified at \SI{1.14(1)}{\at} attributed to residual organic molecules used for exfoliation. The distribution of hydrogen within the material was not analysed, because the experimental analysis conditions were not specifically optimized for hydrogen detection~\cite{2024Gau}.

\begin{figure}[!htb]
   \centering
    \includegraphics[width=\textwidth]{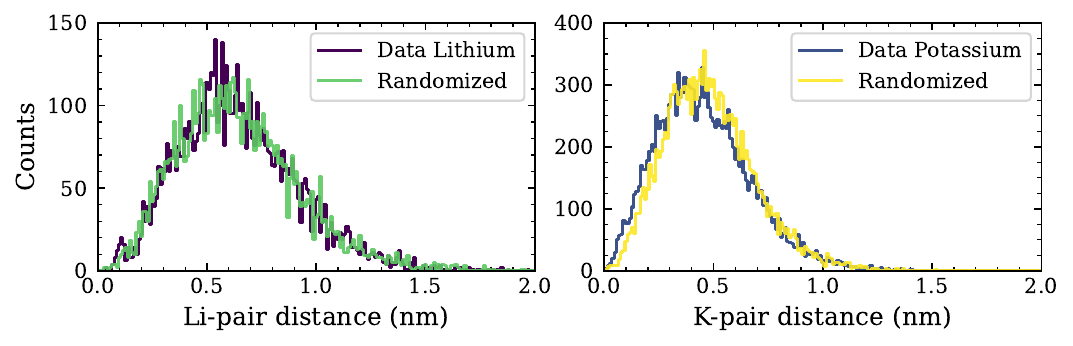}
   \caption{First nearest-neighbour analysis for lithium and potassium in 2D \ce{Ti_{0.87}O2}, calculated within a sub-volume containing more than \SI{80}{\at} \ce{Ti} + \ce{O}. Sample width ion-pair \SI{0.01}{\nano\meter}.}
    \label{fig:Cluster}
\end{figure}

A nearest-neighbour analysis~\cite{2007Ste} was performed to investigate whether the distribution of alkali metals deviates from a randomized distribution, indicating a potential tendency toward clustering. Analysis was restricted to regions containing Ti + O concentrations over \SI{80}{\at} to minimize artifacts such as intermixing zones in the reconstruction, introduced by trajectory aberrations and variable magnification due to differences in the evaporation field of the palladium coating and the 2D material~\cite{1987Mil}. While no deviation from the randomized first nearest-neighbour distribution is observed for lithium in \autoref{fig:Cluster}, a deviation is apparent for potassium. Both results are supported by a frequency distribution analysis, in which the Pearson coefficient $\mu$ is calculated as the normalized measure of a chi-squared statistical test~\cite{2008Moo}. A Pearson coefficient of 0 denotes a completely random distribution, whereas a value of 1 denotes a perfect association, with all intermediate values reflecting the degree of clustering. For lithium, $\mu_{Li}$ equals \num{0.064}, indicating a homogeneous distribution, whereas for potassium, $\mu_{K}$ is \num{0.392}, suggesting a tendency toward cluster formation.

During the cation exchange reaction used to synthesize the 2D material, the alkali metal ions are replaced by protons through treatment in an acid solution (\SI{1}{\Molar} \ce{HCl}) for several days (see Experimental Section for more details). Even though most of the two alkali metal cations could be exchanged, their remaining presence raises concerns about their possible impact as unavoidable dopants in the final material, which may be overlooked during characterization due to concentrations close or below the detection limits of XPS or electron spectroscopy techniques. The homogeneous distribution of lithium suggests that the retention is not governed by diffusion or kinetic barriers, but rather by thermodynamic stabilization at defects or surface sites. Localized electric fields at vacancies or surface wrinkles may cause alkali metal cations to migrate towards these defects. Such stabilisation or adsorption of alkali metals at defects or the surface has been observed for 2D \ce{MoS2}~\cite{2020Kim} and 2D MXenes~\cite{2024Kra_AM, 2024Wya}, suggesting that this behaviour may be a fundamentally characteristic of 2D materials. Protons may likewise become incorporated within defect structures, even though their quantification is not feasible under the present experimental conditions. Potassium clustering could result from a cation size-dependent effect, since not all defects are equally available for stabilization by the larger potassium ions. Sodium was similarly found to form clusters within 2D \ce{MoS2}~\cite{2020Kim}. While crystallization of alkali metal halide salts has been reported during the drying of MXene films~\cite{2025Fav}, such salt formation appears unlikely due to the repeated replacement of the \ce{HCl} solution during the cation exchange reaction and the absence of chlorine signals in the APT data. 

\subsection*{Implications for Functional Properties}

To summarize, XPS analysis confirms that the oxidation states in 2D \ce{Ti_{0.87}O2} are identical to those in stoichiometric \ce{TiO2}. APT analyses reveal a deviation from the expected oxygen to titanium ratio, indicating the formation of oxygen vacancies within the 2D TMO. APT further reveals the incorporation of residual lithium and potassium in \ce{Ti_{0.87}O2}, which are expected to be removed during the cation exchange reaction. These observations suggest that lithium removal from the titanium sublattice generates negatively charged, thermodynamically unstable vacancies. Their stabilization likely arises through a reconstruction by the formation of oxygen vacancies and the adsorption of alkali metal cations. This points to a stabilization mechanism that may be generalizable to defects in 2D materials. To a first approximation, the latter can be rationalized from thermodynamics, as the incorporation of impurities during wet chemical synthesis increases the system's mixing entropy~\cite{2025Gau}. However, due to compositional biases inherent in the APT analysis of oxides, precisely quantifying the deviation of the oxygen content from the nominal composition of the precursor remains extremely challenging.

To consider \ce{Ti_{0.87}O2} for commercial applications, it is essential to understand how defects such as vacancies and adsorbed atoms introduced during synthesis affect the material's electronic structure, and thus functional properties. Experimental and atomistic simulation studies on 2D \ce{Ti_{0.87}O2} reveal that oxygen vacancies can substantially narrow the band gap compared to the defect-free structure~\cite{2018Uch, 2015Son}. Enhancement of dielectric functionality has been achieved through niobium doping, which introduces beneficial lattice distortions~\cite{2011Osa}. In \ce{Ti_{0.87}O2}, the formation of oxygen vacancies and the stabilization of alkali metal cations could counterbalance each other, thereby preserving the exceptional high-$\kappa$ dielectric functionality despite the presence of oxygen vacancies. As such, this mechanism could offer a strategic way to tailor and optimise the dielectric functionality of the material. However, comparing reported functional properties remains difficult without a clear understanding of the precise level of dopants in the material~\cite{2020Lim}. Using 2D TMOs as a template for further synthesis~\cite{2017Sek, 2023Hag}, without detailed compositional characterization, may lead to misleading conclusions about structure–property relationships, as alkali metals are likely introduced as dopants, potentially unnoticed and unaccounted for thus so far.

\section*{Conclusion}

Advances in specimen preparation and analytical techniques continue to push the boundaries of compositional analysis of 2D materials. For 2D TMO \ce{Ti_{0.87}O2}, APT revealed both oxygen loss compared to the bulk \ce{K_{0.8}[Ti_{1.73}Li_{0.27}]O4} precursor and the presence of lithium and potassium within the material. Both observations are believed to contribute to the stabilization of negatively charged vacancies on the titanium sublattice, formed during the cation exchange reaction used to synthesize the 2D material. Successful integration in nanoelectronics requires a deep understanding of the interactions between oxygen vacancy formation and stabilization of alkali metals on functional properties and device performance. Further investigation is required to control the density of these defects as potential dopants, which could enhance functionality, potentially by optimizing synthesis conditions like reaction time and chemical concentration during the cation exchange process. Given the detrimental effect of oxygen vacancies on dielectric performance, intentionally retaining alkali metals within the material may provide an effective strategy to mitigate their formation. Such observations bring into question whether the top-down synthesis of these materials is really always as straightforward as expected, or with the level of control of the composition necessary for widespread application.

\section*{Experimental Methods}

\subsection*{Synthesis of Lepidocrocite-Type \ce{K_{0.8}Ti_{1.73}Li_{0.27}O4} Potassium Lithium Titanate}

Stoichiometric amounts of lithium carbonate (\ce{Li2CO3}, Sigma-Aldrich), potassium carbonate (\ce{K2CO3}, Sigma-Aldrich), and anatase-phase titanium dioxide (\ce{TiO2}, Sigma-Aldrich) were weighed in a molar ratio of \num{0.14}:\num{0.40}:\num{1.73}. The powders were mechanically blended in a tumbler with zirconia balls at \SI{180}{\rpm} for \SI{18}{\hour}. After homogenization, the powder mixture was transferred to an alumina crucible and calcined at \SI{1000}{\degreeCelsius} for \SI{20}{\hour} to obtain the layered titanate with the nominal stoichiometry \ce{K_{0.8}[Ti_{1.73}Li_{0.27}]O4}.

\subsection*{Synthesis of 2D \ce{Ti_{0.87}O2} Transition Metal Oxide}

2D \ce{Ti_{0.87}O2} was exfoliated from bulk \ce{K_{0.8}[Ti_{1.73}Li_{0.27}]O4} powder adapting the synthesis protocol reported by Tanaka et al.~\cite{2003Tan}. Initially, \SI{2}{\gram} of \ce{K_{0.8}[Ti_{1.73}Li_{0.27}]O4} was immersed in \SI{200}{\milli\litre} of \SI{1}{\Molar} \ce{HCl} solution and stirred for \SI{120}{\hour} to obtain fully protonated titanate \ce{H_{1.07}Ti_{1.73}O4*H2O}. The \ce{HCl} solution was replaced every \SI{24}{\hour} to ensure complete cation exchange. After protonation, the resulting \ce{H_{1.07}Ti_{1.73}O4*H2O} was added to a tetramethylammonium hydroxide (TMAOH) solution at a H:TMA molar ratio of \num{1}:\num{1} and stirred for \SI{72}{\hour}. The intercalated powder was washed with ethanol and dried under ambient conditions. For exfoliation, the intercalated powder was dispersed in deionized water at a concentration of \SI{12}{\gram\per\litre} and sonicated for \SI{1}{\hour}. The 2D colloidal solution was separated from the unexfoliated precipitates by centrifugation. 

\subsection*{Electron Microscopy}

Secondary electron imaging of the precursor powder and the 2D material film was performed with a Zeiss Sigma 500 SEM. The microscope was operated at \SI{5}{\kilo\volt} and \SI{1.4}{\nano\ampere} with a working distance of \SI{5}{\milli\meter}. SEM imaging of \ce{Ti_{0.87}O2} monolayer nanosheets was performed by spin-coating a \SI{5}{\gram\per\litre} colloidal solution onto an \qtyproduct{8 x 8}{\mm} silicon die at \SI{2000}{\rpm}.

\subsection*{X-ray diffraction}

X-ray diffraction measurements were performed using a Malvern Panalytical Aeris diffractometer over a $2\Theta$ range of \SIrange{3}{85}{\degree} with \SI{0.02}{\degree} steps.

\subsection*{X-ray photoelectron spectroscopy}

XPS spectra were acquired with a Physical Electronics PHI Quantera II spectrometer using an \ce{Al}-K$\alpha$ source at \SI{1486.6}{\electronvolt}. Pass energy was set at \SI{55}{\electronvolt} to record the core level spectra of titanium (\ce{Ti} $2p$), and at \SI{26}{\electronvolt} to record the core level spectra of oxygen (\ce{O} $1s$) and potassium (\ce{K} $2p$). The survey scan spectrum was obtained with a pass energy of \SI{140}{\electronvolt}. All the measurements were recorded using a take-off angle of \SI{45}{\degree}. Energy step size of \SI{0.025}{\electronvolt} was used for the high-resolution core level spectra, and \SI{0.25}{\electronvolt} was used to collect the survey scan spectrum. No sputter cleaning was performed prior to the measurements, as argon ion sputtering is known to alter the oxidation states of TMOs~\cite{2002Has}.

Peak fitting of the core spectra obtained during the XPS studies was performed using CasaXPS software version 2.3.25. A Shirley-type background was used for all the analysed spectra. Gaussian (\SI{70}{\percent}) - Lorentzian (\SI{30}{\percent}) peak shape was used for all symmetric components. Binding energies were calibrated by referencing the \ce{O} $1s$ core level signal assigned to lattice oxygen at \SI{530.0}{\electronvolt}. The \ce{Ti} $2p$ core level spectrum was fitted with two components corresponding to Ti(IV), with peaks at \SI{\sim 458.4}{\electronvolt} ($2p_{3/2}$) and \SI{\sim 464.1}{\electronvolt} ($2p_{1/2}$). The \ce{O} $1s$ core level spectrum was fitted using a lattice oxygen peak at \SI{\sim 530.0}{\electronvolt} and a peak at \SI{\sim 532.2}{\electronvolt}, which likely arises from surface adsorbed species and adventitious organic contamination. Two components were fitted in the \ce{K} $2p$ core level spectrum, corresponding to the $2p_{3/2}$ and $2p_{1/2}$ peaks at \SI{\sim292.6}{\electronvolt} and \SI{\sim 295.5}{\electronvolt}, respectively

\subsection*{Atom Probe Tomography}

Needle-shaped APT specimens were prepared using a scanning electron microscope coupled with a gallium focused ion beam (Helios Nanolab 600, FEI) following the workflow introduced for 2D material films by Krämer et al.~\cite{2024Kra_MAM}. This workflow involves sputtering of a thin conformal metallic coating of the fragile APT specimen to stabilize it mechanically~\cite{2024Sch}, and to suppress preferential alkali metal migration due to the intense electrostatic field during APT analysis~\cite{2024Sin, 2025Yad}. Palladium (EVOCHEM Advanced Materials, \SI{99.95}{\percent} purity) was chosen as a chemically inert sputtering material to reduce incorporation of oxygen during the coating procedure (see Supporting Information for more details). Sputtering parameters for palladium were \SI{30}{\kilo\volt} and \SI{48}{\pico\ampere} for \SIrange{45}{60}{\second}, repeated \num{4} times after rotating the specimen by \SI{90}{\degree}.

APT analyses of the 2D TMO were performed using a 5000XS (straight flight path) local electrode atom probe (Cameca Instruments), operating in ultraviolet laser-pulsing mode. Parameters were set to a base temperature of \SI{60}{\kelvin}, a laser pulse energy of \SI{50}{\pico\joule}, a laser pulsing rate of \SI{125}{\kilo\hertz}, and a target detection rate of \num{5} ions per \num{1000} pulses on average. For the calibration measurements of the bulk \ce{K_{0.8}[Ti_{1.73}Li_{0.27}]O4} powder, a 5000XR (reflectron-fitted) local electrode atom probe (Cameca Instruments) was used. The analysis parameters were identical to those used for the 2D material, except that the laser pulse energy was varied between \SI{1}{\pico\joule} and \SI{50}{\pico\joule} to create different electrostatic field conditions during field evaporation.

Data reconstruction and analysis were done with AP Suite 6.3 by Cameca Instruments following the default voltage-based reconstruction algorithm.

\FloatBarrier

\section*{Supporting Information Available}

Optimisation of the In Situ Coating Procedure for Atom Probe Specimen Preparation of 2D Transition Metal Oxides; Additional experimental APT and XPS data


\section*{Author Contributions}

M.K. and B.G. conceived and conceptualized the study. B.F. synthesized the material. M.K. carried out the APT analysis, and J.M.P. performed the XPS analysis. M.K. and B.G. led the data analysis and interpretation, with support from B.F., J.M.P., and A.A.. B.A.R., N.E., M.S., and B.G. provided funding, resources, and supervision. M.K. and B.G. drafted the manuscript with input from all authors. All authors contributed to the discussion of the results and agreed on the conclusions.


\section*{Acknowledgment}

The authors acknowledge financial support from the German Research Foundation (DFG) through DIP Project No. 450800666. The support to the SEM, FIB, and APT facilities at MPI SusMat by Katja Angenendt, Christian Broß, Andreas Sturm, and Uwe Tezins is gratefully acknowledged.


\section*{Declaration of Generative AI and AI-assisted technologies in the writing process}

During the preparation of this work, the authors used DeepL in order to improve language and readability. After using this tool, the authors reviewed and edited the content as needed and take full responsibility for the content of the publication.


\section*{Conflict of Interest}

The authors declare no conflict of interest.


\section*{Data Availability Statement}

The data that support the findings of this study are available from the corresponding authors upon reasonable request.

\printbibliography

\end{document}